\documentclass{aa}

\usepackage{graphics}
\usepackage{longtable}
\usepackage{txfonts}
\usepackage{graphicx}
\usepackage{hyperref}
\usepackage{multirow}
\usepackage{supertabular}

\begin{document}

\title{The binary Be star $\delta$ Sco at high spectral and spatial resolution\thanks{Based on observations made with ESO Telescopes at Paranal under programs 383.D-0210, 385.D-0275, and CHARA/VEGA observations}}
\subtitle{Disk geometry and kinematics before the 2011 periastron}
\authorrunning{Meilland et al. }
\titlerunning{The binary Be star $\delta$ Sco at high spectral and spatial resolution: I. before the 2011 periastron}

\author{A. Meilland \inst{1}, O. Delaa \inst{2}, Ph. Stee \inst{2}, S. Kanaan\inst{3}, F. Millour \inst{2}, D. Mourard \inst{2}, D. Bonneau\inst{2}, R. Petrov \inst{2},  N. Nardetto \inst{2}, A. Marcotto \inst{2}, A. Roussel \inst{2}, J.M. Clausse \inst{2}, K. Perraut \inst{4}, H. McAlister\inst{5,6}, T. ten~Brummelaar\inst{6}, J. Sturmann\inst{6}, L. Sturmann\inst{6}, N. Turner\inst{6}, S. T. Ridgway\inst{7}, C. Farrington\inst{6} and P.J. Goldfinger\inst{6}}

\offprints{meilland@mpifr-bonn.mpg.de}

\institute{
Max Planck Intitut fur Radioastronomie, Auf dem Hugel 69, 53121 Bonn, Germany 
\and
UMR 6525 CNRS H. FIZEAU  UNS, OCA, CNRS, Campus Valrose, F-06108 Nice cedex 2, France.
\and
 Departamento de F\'isica y Astronom\'ia, Universidad de Valpara\'iso, Chile.
\and
UJF/CNRS LAOG, 414, rue de la Piscine, Domaine Universitaire 38400 Saint-Martin d'H\`eres, France
\and
Georgia State University, P.O. Box 3969, Atlanta GA 30302-3969, USA
\and
CHARA Array, Mount Wilson Observatory, 91023 Mount Wilson CA, USA
\and
National Optical Astronomy Observatory, P.O. Box 26732, Tucson, AZ 85726-6732, USA}

   \date{Received; accepted }

   \abstract
{Classical Be stars are hot non-supergiant stars surrounded by a gaseous circumstellar disk that is responsible for the
observed IR-excess and emission lines. The influence of binarity on these phenomena remains controversial. }
{$\delta$ Sco is a binary system whose primary suddently began to exhibit the Be phenomenon at the last periastron in 2000. We want to constrain the geometry and kinematics of its circumstellar environment.}
{We observed the star between 2007 and 2010 using spectrally-resolved interferometry with the VLTI/AMBER and CHARA/VEGA instruments.}
{We found orbital elements that are compatible with previous estimates. The next periastron should take place around July 5, 2011 ($\pm 4$\,days). We resolved the circumstellar disk in the H$\alpha$ (FWHM $= 4.8\pm1.5$\,mas), Br$\gamma$ (FWHM $= 2.9\pm 0.5$\,mas), and the 2.06$ \mu$m He\,{\sc i} (FWHM $= 2.4\pm 0.3$\,mas) lines as well as in the K band continuum (FWHM $\approx 2.4$\,mas). The disk kinematics are dominated by the rotation, with a disk expansion velocity on the order of 0.2\,km\,s$^{-1}$. The rotation law within the disk is compatible with Keplerian rotation.}
{As the star probably rotates at about 70$\%$ of its critical velocity the ejection of matter doesn't seems to be dominated by rotation. However, the disk geometry and kinematics are similar to that of the previously studied quasi-critically rotating Be stars, namely $\alpha$ Ara, $\psi$ Per and 48 Per.}

   \maketitle
%

\section{Introduction} 

\begin{table*}[!t]
\caption{\label{log}VLTI/AMBER and CHARA/VEGA observing logs for $\delta$ Sco.}
\centering \begin{tabular}{@{~}cccccccc@{~}}
\hline
\multicolumn{8}{c}{VLTI/AMBER}\\
\hline
Observing Time &Telescopes&Base Length  &Position Angle &Mode&Exposure/Frame&Seeing&Calibrators\\
Start (UTC)&&(m)&($^o$)&&(s)&(")&(HD)\\
\hline\hline
2007-09-05 23:52&D0-H0-G1&67/56/71&-28/82/20&MR-K-F&1.00&0.67&146791\\
2008-05-03 06:12&A0-K0-G1&90/90/127&-151/-61/-106&LR-HK&0.05&0.79&139663\\
2008-05-24 04:28&A0-D0-H0&64/32/96&-109/-109/-109&LR-HK&0.05&0.99&139663, 132150\\
2008-05-25 04:05&A0-D0-H0&64/32/96&-110/-110/-110&LR-HK&0.05&0.81&139663\\
2008-07-11 01:48&A0-K0-G1&90/90/127&-151/-61/-106&LR-HK&0.05&0.97&139663, 166295\\
2009-04-11 07:40&A0-K0-G1&90/90/128&-151/-62/-106&LR-HK-F&0.05&0.63&139663\\
2009-04-11 08:19&A0-K0-G1&90/88/125&-148/-57/-103&LR-HK-F&0.05&0.82&139663\\
2009-06-04 05:20&E0-G0-H0&30/15/45&-100/-100/-100&LR-HK-F&0.05&0.74&139663\\
2009-06-04 05:20&E0-G0-H0&28/14/42&-97/-97-/97&LR-HK-F&0.05&0.94&139663\\
2009-07-30 02:27&D0-H0-G1&67/55/71&-26/83/20&MR-K-F&1.00&0.76&139663\\
2010-04-15 07:12&D0-H0-G1&64/71/71&72/-172/135&MR-K-F&0.50&0.82&139663\\
2010-04-15 07:55&D0-H0-G1&63/71/71&76/-167/140&MR-K-F&0.50&0.94&139663\\
2010-04-19 05:14&D0-H0-G1&60/71/69&60/177/127&HR-K-F (2.17$\mu$m)&6.00&0.85&139663\\
2010-04-20 09:14&D0-H0-G1&54/71/66&84/-159/155&HR-K-F (2.06$\mu$m)&6.00&0.60&139663\\
2010-05-10 05:11&D0-H0-G1&128/90/90&-110/115/-156&LR-K&0.05&0.83&139663\\
2010-05-10 05:52&D0-H0-G1&126/90/90&-107/119/-90&LR-K&0.05&0.95&139663\\\hline\hline

\multicolumn{8}{c}{CHARA/VEGA}\\
\hline
Observing Time&Telescopes&Length&Position Angle&Mode&Detector&r$_{0}$&Calibrators\\
Start (UTC)&&(m)&P($^o$)&&&(cm)&(HD)\\
\hline\hline
2010-05-05 08:50&S1-S2&20&-16 &MR&Red& 17&144470\\
2010-05-05 09:37&S1-S2&21&-25 &MR&Red&18&144470\\
2010-05-05 10:14&S1-S2&23&-30 &MR&Red& 17&144470\\
2010-06-24 04:31&E1-E2&63&-111&MR&Red& 9&144470\\\hline 
\hline

\end{tabular}

\end{table*}

\begin{figure*}[!htb]
       \centering
       \hspace{-0.5cm}  
       \includegraphics[width=0.48\textwidth]{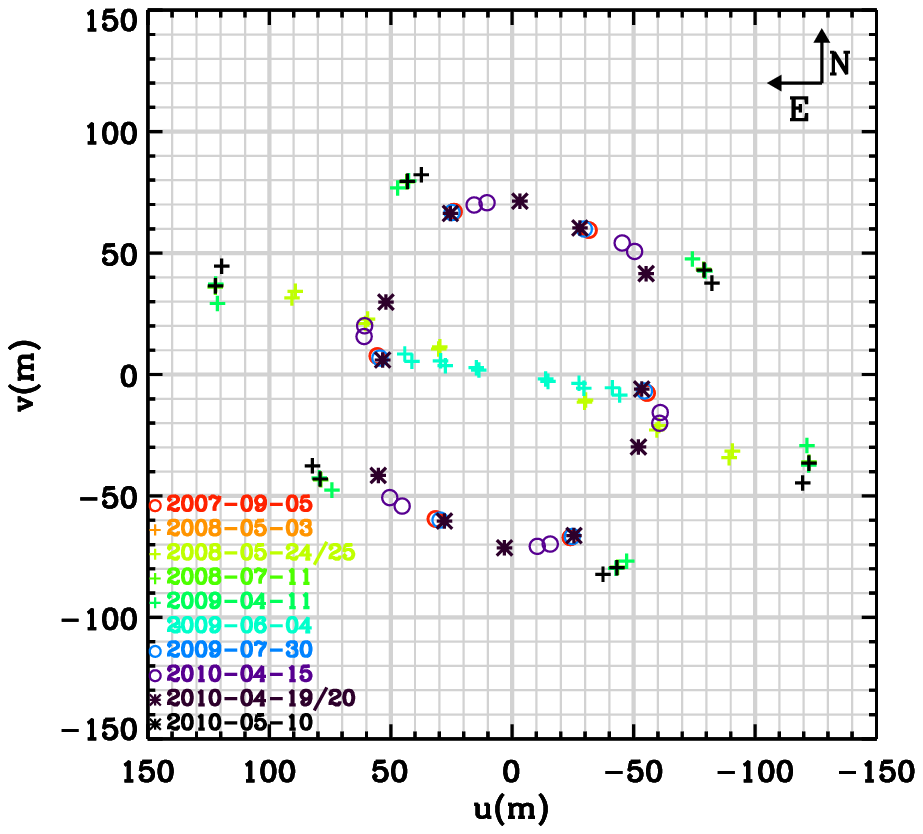}
       \hspace{1.0cm}
       \includegraphics[width=0.48\textwidth]{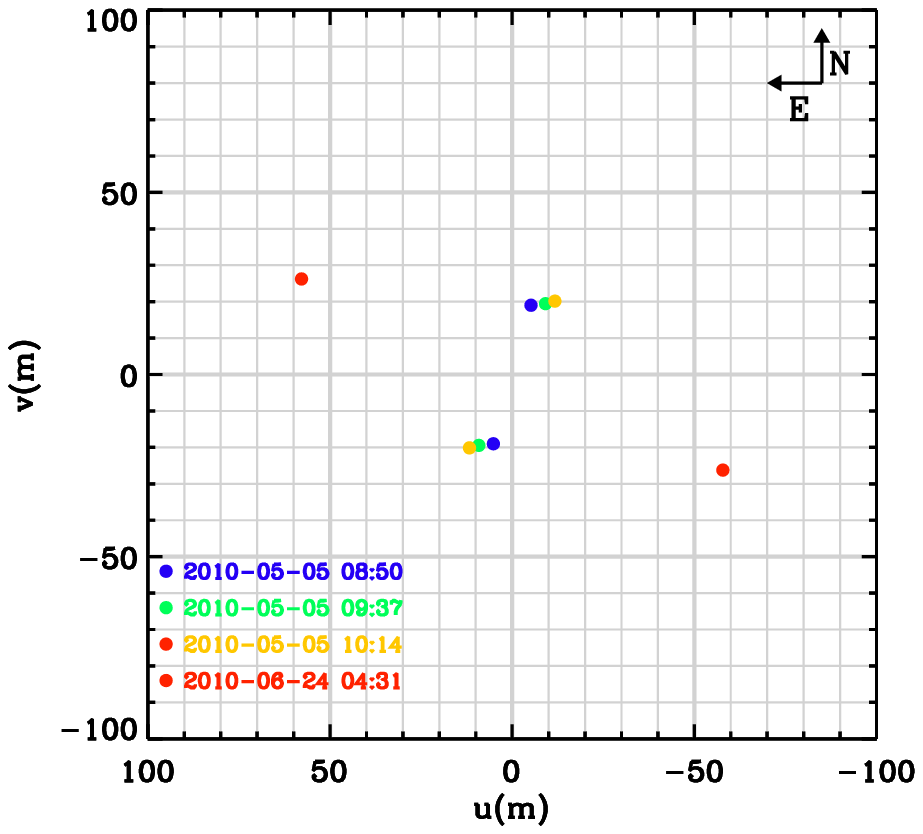}
\caption{(u,v) coverage for the complete VLTI/AMBER (left) and CHARA/VEGA (right) 2007-2010 dataset.}
\label{uvplan}
\end{figure*}

Classical Be stars are close-to-main-sequence hot stars that show or have shown infrared-excess and emission lines in their spectra. Such features stem from a dense gaseous circumstellar environment. However, the presence of a diluted wind with a terminal velocity of several hundreds of km\,s$^{-1}$ have also been deduced from ultraviolet spectroscopy (Marlborough \& Peters 1986). Consequently, taking into account the fact that they are also fast rotators, a generally accepted view of these objects is that their circumstellar environment consists of two distinct regions: a dense equatorial disk dominated by rotation where most of the infrared radiation and emission lines are produced and a more diluted polar wind responsible for the highly broadened ultraviolet lines.

There is general agreement on the importance of stellar rotation on the ejection of matter and the break of the spherical symmetry of the circumstellar environment. However, the question of whether or not additional physical processes are needed to produce ejection remains uncertain. Taking into account effects of gravitational darkening, Fr\'emat et al. (2005) derived a mean value for the average rotational velocity of Be stars on the order of 88\,\% of their critical velocity (V$\mathrm{c}$) thus insufficient to fully explain the ejection of matter. In a statistical study of the rotation velocity of 462 Be stars Cranmer (2005) found that early type Be stars (O7-B3) exhibit a roughly uniform spread of intrinsic rotation speed extending from 40\,\% up to 100\,\% $V_\mathrm{crit}$, whereas late types (B3-A0) are all quasi-critical rotators. This is a strong clue that rotation is essential in the formation of circumstellar envelopes for late type Be stars, whereas other physical mechanisms such as radiative pressure, pulsations or magnetism may dominate the ejection of matter for some of the earlier spectral type Be stars.

First VLTI/AMBER observations of Be stars $\alpha$ Ara (Meilland et al. 2007a) and $\kappa$ CMa (Meilland et al. 2007b) have shown evidence of different geometries and kinematics, reinforcing the hypothesis of the heterogeneity of this group of stars in terms of mass-ejection processes (Stee \& Meilland 2009). Moreover, recent discoveries of a companion around Achernar ($\alpha$ Eri, Kervella et al. 2008) and $\delta$ Cen (Meilland et al. 2008) may indicate that the putative effect of binarity on the Be phenomenon might have been underestimated. In the case of Achernar, Kanaan et al. (2008) successfully modeled spectroscopic variations as a brief equatorial outburst propagating into the circumstellar environment. Such phenomena probably originate from the close encounter between the central star and its companion. 

In this context, a detailed study of $\delta$ Sco (HD 143275, HIP 78401) brings new perspectives to the understanding of the Be phenomenon. This bright southern object has long been studied and first evidence of its multiplicity was reported by Innes (1901) using the lunar occultation technique. However, this work was forgotten for a long time, and the binary nature of $\delta$ Sco was rediscovered with three different techniques in 1974: by speckle-interferometry (Labeyrie et al. 1974), lunar occultation (Dunham 1974), and Intensity interferometry (Hanbury Brown et al. 1974). Using various interferometric measurements Bedding (1993) deduced an orbit of 10.6\,years. However, the $\delta$ Sco system did not show clear evience of the Be phenomenon until the last periastron in June 2000. At this epoch, Otero (2001) found a  0.4\,mag brightening of the object. Simultaneous spectroscopic observations published in Fabregat et al. (2000) showed evidence of strong H$\alpha$ emission lines.

\begin{table*}[!t]
\caption{2007-2010 evolution of the $\delta$ Sco binary system determined from our VLTI/AMBER observations.\label{binfit}}
{\centering \begin{tabular}{c|cccc|c|cc}
\hline

Date&\multicolumn{4}{c}{Model parameters}\vline&&\multicolumn{2}{c}{Separation in polar coordinates}\\

							&\textit{x} 						&\textit{y} 			 				&\textit{F$_1$}					&\textit{D$_1$}				&$\chi^2_r$ 			&\textit{sep} 			 	&$\theta$\\
							&(mas)				&(mas)					&								& (mas)				& 			&(mas)			& (deg)\\
\hline\hline
2007-09-05		&-3.9$\pm$0.3	&-168.9$\pm$0.3	&0.96$\pm$0.06	&1.7$\pm$0.4	&2.8		&168.9		 	&1.3\\
2008-05-03		&-						&-							&1.00						&1.7$\pm$0.6	&10			&-					&-\\
2008-05-24/25	&-7.8$\pm$0.4	&-163.5$\pm$1.0	&0.96$\pm$0.01	&1.4$\pm$0.1	&6.1		&163.7			&2.7\\
2008-07-11		&-6.5$\pm$0.4	&-163.0$\pm$0.5	&0.99$\pm$0.03	&0.9$\pm$0.1	&1.2		&163.1			&2.3\\
2009-04-11		&-11.5$\pm$1.0&-156.8$\pm$1.0	&0.98$\pm$0.02	&2.0$\pm$0.1	&3.6		&157.2			&4.2\\
2009-06-04		&-12.7$\pm$0.5&-146.8$\pm$3.0	&0.94$\pm$0.01	&1.7$\pm$0.1	&0.9		&147.3			&4.9\\
2009-07-30		&-12.8$\pm$1.5&-135.3$\pm$2.5	&0.93$\pm$0.01	&2.7$\pm$0.1	&2.3		&135.9			&5.4\\
2010-04-15		&-27.7$\pm$0.1&-101.6$\pm$0.1	&0.94$\pm$0.02	&1.9$\pm$0.5	&2.9		&105.3			&15.3\\
2010-05-10		&-22.6$\pm$0.1&-96.3$\pm$0.1	&0.95$\pm$0.1		&1.4$\pm$0.1	&2.4		&98.9				&13.2\\
\hline\hline
\end{tabular}\par}
\end{table*}

In this work, we present near-infrared VLTI/AMBER (Petrov et al. 2007) and visible CHARA/VEGA (Mourard et al. 2009) interferometric measurements of $\delta$ Sco. The paper is organized as follows. In Sect. 2 we present the observations and the data reduction process. In Sect. 3 modeling of the data in the continuum allows us to constrain the binary orbit and physical parameters. In Sect. 4 we model the emission lines and constrain the envelope geometry and kinematics. Finally a short discussion in Sect 5 is followed by conclusions in Sect. 6.

\section{Observation and data reduction}

\subsection{VLTI/AMBER}
We initiated an interferometric follow-up of $\delta$ Sco soon after the opening of the VLTI/AMBER instrument to the scientific community. The star was then regularly observed between 2007 and 2010. The corresponding observing log is given in Table~\ref{log}. Since this target is bright enough, i.e $m_H \approx m_K \approx 2.4$, all observations were carried out using the 1.6 m auxiliary telescopes (AT). Low Resolution ($R = 30$), Medium Resolution ($R = 1500$), and High Resolution ($R = 12000$) spectral modes were used during the campaigns. The VLTI fringe tracker FINITO was used for all observations except those in 2008.

The data were reduced using the AMBER data reduction software \texttt{amdlib}, version 2.2. (Tatulli et al. 2007). The average raw complex visibility and closure phase was determined using the standard method, keeping the 20\% of the frames with the higher SNR ratio. The interferometric calibration was then done using custom scripts described in Millour et  al. (2008)\footnote{The scripts are available to the community at the following webpage: \url{http://www.mpifr-bonn.mpg.de/staff/fmillour}}. The (u,v)  coverage for all observations is plotted in Fig~\ref{uvplan}.

\begin{figure}[t]
       \centering  
       \includegraphics[width=0.48\textwidth]{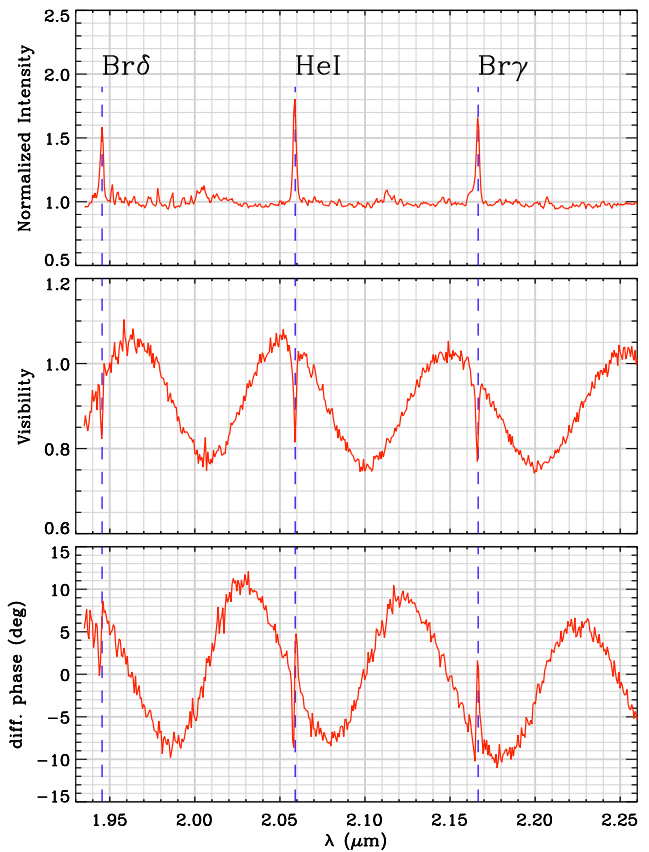}
\caption{Example of a spectrally resolved VLTI/AMBER measurement for one baseline in MR mode. The oscillations of the visibility and phase plotted as the function of the wavelength are due to the binarity of the object. This figure also exhibit effects of the $\delta$ Sco circumstellar environment geometry and kinematics in the Br$\gamma$, He\,{\sc i} and Br$\delta$ emission lines.}
\label{visMR}
\end{figure}

Oscillations due to the binarity of $\delta$ Sco are seen in all LR and MR observations expect those of 2008-05-03.  Moreover three emission lines (Br$\gamma$, He\,{\sc i} 2.06$\mu$m and Br$\delta$)  Fig~\ref{visMR} are clearly visible in the MR observations as shown in Fig.~\ref{visMR}.
        
\subsection{CHARA/VEGA}

$\delta$ Sco ($m_V\approx2.3$) was observed with the CHARA interferometer (ten Brummelaar et al. 2005) using the newly available VEGA instrument (Mourard et al. 2009). This 2-4 telescope visible beam-combiner has currently the highest spatial resolution ($\theta_{\rm min} = 0.3$\,mas at 0.4\,$\mu$m with the 330m baseline) and highest spectral resolution for an optical/IR interferometer  (up to $R = 30000$ ). Observations were carried out in May and June 2010 using the MR mode centered on the H$\alpha$ line (656.3nm). We obtained four measurements, three with the  S1-S2 baseline ($\approx 30$\,m), and one with the WI-W2 baseline ($\approx 70$\,m). The log of the observations including the star used as an interferometric calibrator are shown in Table~\ref{log}.

The data were reduced using the standard CHARA/VEGA reduction software developed by the VEGA team and described in Mourard et al. (2009). This software offers two data reduction modes, the first based on spectral density analysis and the second on the cross-spectral scheme.

\section{The binary system}

\subsection{Calculating the binary separation}
\label{litprofit}

We used the \texttt{LITpro}\footnote{LITpro software available at http://www.jmmc.fr/litpro} 
 model fitting software for optical/infrared interferometric data developed by the Jean-Marie Mariotti Center (JMMC) to analyze our data. It is based on the Levenberg-Marquardt algorithm, that allows from a set of initial values of the model parameters to converge to the closet $\chi^2$ local minimum. It also include tools to facilitate the search for the global minimum.
{\bf LITpro calculates an error on the fitted parameters based on the $\chi^2$ value at the minimum. It uses data error estimates based on the OI FITS format, which does not include error correlation estimates. Therefore, in some cases, LITpro can provide underestimated errors on the parameters.}

Since the aim of this first study is to refine the previous determination of the binary orbit, we only used AMBER LR-K and MR-K data, where the binary signal is clearly visible. In our analysis, we also neglect signal coming from the K band emissions lines visible in Fig~\ref{visMR}.
 
The primary is modeled as a uniform disk to take into account partial resolution of its circumstellar environment in the continuum, whereas the companion is modeled as a point like source. Consequently our model has only 4 free parameters: the binary separation in Cartesian coordinates  ($x$ and $y$) where $x$ (resp. $y$) is counted positive toward the east (resp. north), the relative flux of the primary to the total flux (\textit{F$_1$}), and its diameter (\textit{D$_1$}). 

The result of the model-fitting is presented in Table~\ref{binfit}. The binary separation decreased from about 170 mas in 2007 to 100 mas in 2010. 
{\bf The estimated error from LITpro, in average 0.8\,mas on x and y, is reported in this table. The data points scatter around the fitted orbit (Sect. \ref{orbitelement}) is of the order of 7 mas. Therefore, our error estimates are very likely underestimated.}
The K-band primary relative flux $F_1$ is varying from 92 to 100\,$\%$ of the total flux. However, this variation is linked to the spatial frequency sampling since it depends on the baseline length and spectral resolution. The minimum values of $F_1$ are obtained for the MR observations. For the LR ones its value varies from 98$\pm$1$\%$ for observations made with the longest triplet A0-K0-G1 to 94\,$\%$ for the shorter triplet E0-GO-H0. By studying the measurements with the intermediate triplet A0-D0-H0, we clearly see a change in the oscillation amplitudes from the short baselines to the largest. Such a variation could have been a consequence of resolving the companion. However, since these oscillation amplitudes determined using observations made in MR and LR modes for the same baseline length are mutually inconsistent, this clearly indicates that it comes from a technical problem due to the sampling of oscillations. Finally, we can deduce that the amplitudes measured in MR mode are not biased by this spatial frequency sampling effect. Thus the real K-band relative flux of the primary is on the order of 93\,$\pm 1 \%$.
 
According to almost all measurements, the primary component is partially resolved. Modeling it as a simple uniform disc we can derive a diameter of 1.6\,$\pm 0.6$ mas.  Adopting the stellar radius from Carciofi et al. (2006), 7\,$R_\odot$, and considering a distance of 150\,pc (van Leeuwen 2007), this corresponds to 25.7\,$\pm9.6 R_\odot$, i.e. 3.67\,$R_\star$. This gives us a lower limit of the circumstellar disk extension in the K band for the case where it fully dominates the primary K band flux. If only half of the primary K band emission would originate from the disk, as for $\alpha$ Ara (Meilland et al. 2007a) or $\kappa$ CMa (Meilland et al. 2007b), its extension would be on the order of 2.4\,mas, i.e. 4.6\,$R_\star$. 
\begin{figure*}[!t]
       \centering  
       \includegraphics[width=0.85\textwidth]{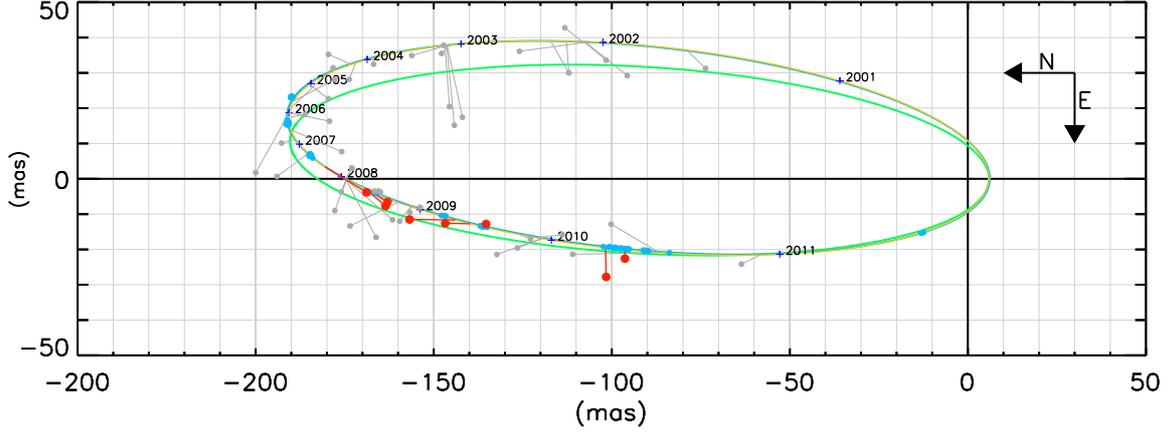}
\caption{The visual orbit of $\delta$ Sco. The gray circles are binary separations taken from the \textit{Fourth Catalog of Interferometric Measurements of Binary Star}, the blue ones from Tycner et al. (2011) and the red ones are our new measurements derived from our VLTI/AMBER data. The green and blue solid lines represent the orbits derived by Tango et al. (2009) and Tycner et al. (2011), respectively. Our best-fit orbit is plotted as an orange solid line.}
\label{orbit}
\end{figure*}

\begin{figure*}[!t]
       \centering  
       \includegraphics[width=0.85\textwidth]{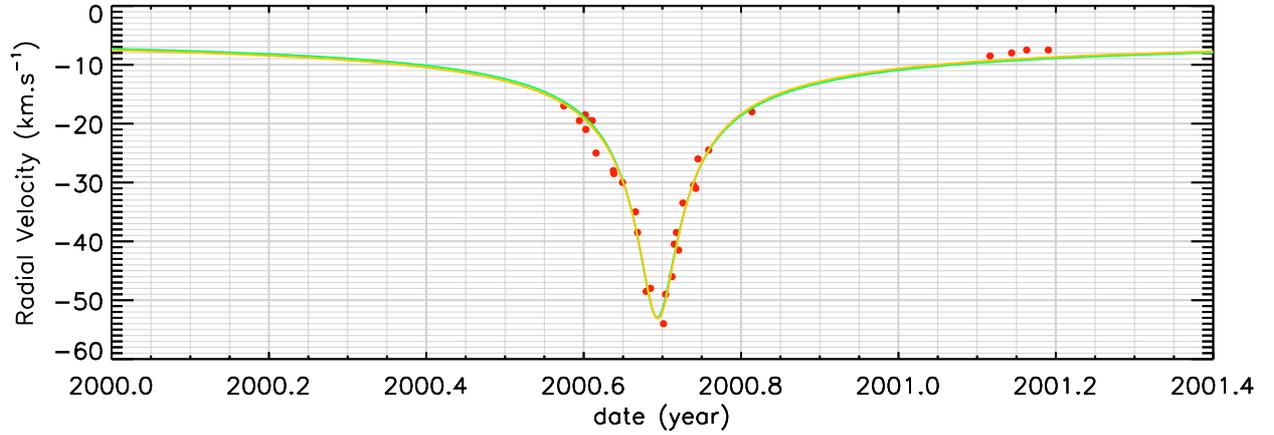}
\caption{$\delta$ Sco radial velocity around the last periastron. Data (red circles) are taken from Miroshnichenko (2001). The green and orange lines represent Tango et al. (2009) and our best-fit model, respectively.}
\label{Rv}
\end{figure*}

\subsection{The orbital elements}
\label{orbitelement}

\subsubsection{Previous work}

Since the rediscovery of its binarity in 1974, $\delta$ Sco has been observed several times using speckle-interferometry, mainly published in McAlister \& Harkopft (1988). A first attempt to constrain its orbit using this set of data as well as optical-interferometry data from the MAPPIT experiment was by Bedding (1993). New speckle-interferometric measurements allowed Hartkopf et al. (1996) to derive more accurate and significantly different parameters. Using the same dataset complemented by radial velocity measurements close to the 2000 periastron, Miroshnichenko et al. (2003) refined the previous analysis. Tango et al. (2009) have used the same dataset but with a more consistent model-fitting algorithm from Pourbaix (1998) to estimate the orbit. Finally the most recent analysis of $\delta$ Sco orbit was by Tycner et al. (2011) using 96 measurments from taken with the Navy Prototype Optical Interferometer (NPOI). The resulting orbital elements determined in these five papers are presented in Table~\ref{orbitalelements}. 

The periastron passage ($T_0$) and the eccentricity are extremely well constrained by the radial-velocity measurements and small uncertainties remain for the orbit angular parameters \textit{i}, $\omega$, and $\Omega$. The main differences between the values of the orbital elements derived by two latest and most complete studies are the orbit period ($P$) and major-axis  (\textit{a}). One important parameter for the ongoing $\delta$ Sco observing campaigns is the exact date of the next periastron. According to Miroshnichenko et al. (2001) it should take  place in 2011 on April 9, whereas Tango et al. (2009) predict that this will occur between May 30 and June 14 and Tycner et al. (2011) that the next periastron passage is expected to occur on UT 2011 July 6\,$\pm 2$\,d.

\subsubsection{This work}

All our measurements are plotted in Fig~\ref{orbit}. They are roughly compatible with the most recent orbits determined by Tango et al. (2009) and Tycner et al. (2011) that are over-plotted on this figure. Other measurements available from the \textit{Fourth Catalog of Interferometric Measurements of Binary Star}$\footnote{http://ad.usno.navy.mil/wds/int4.html}$ are also over-plotted. 

In Cartesian coordinates the $\chi^2$ for the separation measurements is given by: 

\begin{equation}
\chi^2_{\rm sep}(P,T_0,a,e,i,\omega,\Omega) =\sum_{i=1}^{N_{\rm sep}}\Biggl[\Bigl(\frac{x_i-\widehat{x}_i}{\sigma_{x,i}}\Bigr)^2+\Bigl(\frac{y_i-\widehat{y}_i}{\sigma_{y,i}}\Bigr)^2\Biggr]
\end{equation}

where x$_i$ and y$_i$ is the measured binary separation in Cartesian coordinates for the N$_\mathrm{sep}$=150 measurements (i.e, 96 from Tycner et al 2011, 46 from the USNO database, and 8 from our measurements), $\sigma_{x,i}$ and $\sigma_{x,i}$, the  uncertainties of the measurements, $\widehat{x}_i$ and $\widehat{y}_i$ the modeled separation at the corresponding epochs. $\chi^2_{\rm sep}$ is a function of the 7 orbital parameters: $P$, $T_0$, $a$, $e$, $i$, $\omega$, and $\Omega$.

As in Tango et al. (2009) we also use radial velocities from Miroshnichenko et al. (2001) plotted in Fig~\ref{Rv} in our model-fitting process. The $\chi^2$ for these measurements is given by:

\begin{equation}
\chi^2_{v_{\mathrm{rad}}}(T_0,P,e,\omega,V_0,K_A)=\sum_{i=1}^{N_{v_{\mathrm{rad}}}}\Bigl(\frac{{v_{\mathrm{rad}}}_i-\widehat{v_{\mathrm{rad}}}_i}{\sigma_{{v_{\mathrm{rad}}}_i}}\Bigr)^2
\end{equation}

where $v_{\mathrm{rad}},i$ is the measured radial velocity for the $N_{v_{\mathrm{rad}}} = 30$ measurements, $\sigma_{v_{\mathrm{rad}},i}$ the corresponding uncertainty on the measurements, and $\widehat{v_{\mathrm{rad}}}_i$ the modeled radial velocities. $\chi^2_{v_{\mathrm{rad}}}$ is a function of the orbital parameters $T_0$, $P$, $e$, and $\omega$ and of two additional parameters: the systemic radial velocity of the system $V_0$, and the amplitude of the velocity variations $K_A$.

We simultaneously fit the visual separation measurements and the radial velocities by minimizing the function:

\begin{table*}[!tbh]
\caption{The orbital elements of $\delta$ Sco.\label{orbitalelements}}
\centering \begin{tabular}{cccccccc}
\hline
Orbital el.& Bedding 1993				& Hartkopf et al. 1996 & Miroshnichenko & Tango et al. 2009 & Tycner et al. 2011 &This work\\
	& & & et al. 2001& & &\\
\hline\hline
$P$ (yr)							& 10.5   	& 10.59 $\pm$ 0.075		&	10.58									&10.74$\pm$0.02					& 10.817 $\pm$ 0.005& 10.811$\pm$0.01\\
$T_0$ (yr)					& 1979.3 	& 1979.41 $\pm$ 0.14	& 2000.693 $\pm$ 0.008	&2000.69389$\pm$0.00007	&2000.6927 $\pm$ 0.0014&2000.6941$\pm$0.003\\
$a$ (mas)							& 110			&	106.7 $\pm$ 6.7			&	107										&98.3 $\pm$ 1.2					&99.1 $\pm$ 0.1&98.74$\pm$0.07\\	
$e$										& 0.82		&	0.92 $\pm$ 0.02			&	0.94 $\pm$ 0.01				&0.9401 $\pm$ 0.0002		&0.9380 $\pm$ 0.0007&0.9403$\pm$0.0008\\
$i$	($^o$)						& 70			&	48.5 $\pm$ 6.6			&	38 $\pm$ 5						&38 $\pm$ 6							&32.9 $\pm$ 0.2&30.2$\pm$0.7\\
$\omega$($^o$)			&	170			&	24 $\pm$ 13					&	-1 $\pm$ 5						&1.9 $\pm$ 0.1					&2.1 $\pm$ 1.1&0.7$\pm$2.9\\
$\Omega$($^o$)			&	0				&	159.6 $\pm$ 7.6			&	175										&175.2 $\pm$ 0.6				&172.8 $\pm$ 0.9&174.0$\pm$2.5\\
$V_0$	(km\,s$^{-1}$)	&					&											&	~6										&-6.72$\pm$0.05					&-7&-6.7$\pm$0.2\\
$K_A$ (km\,s$^{-1}$)	&					&											&												&23.83$\pm$0.05					&? &23.9$\pm$0.1\\	

\hline\hline
\end{tabular}

\end{table*}  

\begin{equation}
F (P,T_0,a,e,i,\omega,\Omega,V_0,K_A)=\chi^2_{\rm sep}+\chi^2_{v_{\mathrm{rad}}}
\end{equation}

To minimize F we used the \textit{downhill simplex method} by Nelder \& Mead (1965). The starting parameters were set in four different ways. In the three first fitting processes we used the orbital elements from Tango et al. (2009), Miroshnichenko et al (2001), and Tycner et al (2011). Finally, in the last fitting process we used $10^4$ random starting positions. For each starting position the simplex algorithm was allowed to converge until the difference in $\chi^2$ between two consecutive models is less than $\epsilon = 10^{-4}$.

Using these initial guesses the fit converged to the same solution. The best fit model parameters are shown in Table~\ref{orbitalelements} and the corresponding visual orbit and radial velocity curve are over-plotted on Figs~\ref{orbit}~and~\ref{Rv}, respectively. We found values that fully agrees with the latest estimates from Tycner et al.(2011). This is reasonable taking into account that their measurements dominates the sample both in term of number (96 out of 150) and uncertainties. Finally, we can conclude that the next periastron should take place around July 5, 2011 ($\pm$4 days).

\subsection{The binary system physical parameters}
We can constrain the total mass of the $\delta$ Sco binary system using Kepler's third law and our estimation of the binary period and semi major-axis. Since this last parameter is given in angular units (i.e. mas), the binary system distance is also needed for the mass calculation. Using the distance inferred from the Hipparcos parallax  (Perryman et al. 1997), i.e. $d = 123$\,pc, we derived a total mass of $M_{\rm A+B} = 15.2 \pm 5$\,$M_\odot$. On the other hand, using the distance deduced from van Leeuween (2007) revised Hipparcos parallax, i.e. d~=~150pc, we obtained a total mass M$_{\rm A+B}$~=27.7~$\pm$~10~$M_\odot$. The second estimation is in agreement with the total mass determined previously by Tango et al. (2009), i.e. 23\,$\pm 10$\,$M_\odot$.

From physical modeling of the spectral energy distribution, Carciofi et al. (2006) estimated the mass of the primary to be on the order of $M_{\rm A} = 14$\,$M_\odot$. Assuming that both stars have the same age and are currently on the main sequence, and considering the flux ratio measured in the K band, i.e. $F_{B}/F_{A} = 0.07/0.93 = 0.075 \pm 0.012$, the secondary should also be a quite massive hot star. Thus, it seems that $M_{\rm A+B} = 15.2 \pm 5$\,$M_\odot$ is quite unrealistic. On the other hand $M_{\rm A+B} = 27.7 \pm 10$\,$M_\odot$ would implies that both components have approximatively the same mass.

Finally, Tango et al. (2009) give an estimation of the secondary mass of $M_{\rm B} = 8 \pm 3.6$\,$M_\odot$. From the mass ratio they calculated, i.e. $M_{\rm A}/M_{\rm B} = 1.957 \pm 0.011$ we can deduce the Luminosity ratio using Mass-Luminosity relation from Griffiths et al. (1988): 

\begin{equation}
\frac {L_{\rm B}}{L_{\rm A}}=\Bigl(\frac{M_{\rm B}}{M_{\rm A}}\Bigr)^{3.51\pm0.14}
\end{equation}

Thus, we can deduce that $L_{\rm B}/L_{\rm A} = 0.095 \pm 0.01$. This is compatible with our K band flux ratio measurements, i.e. $F_{\rm B}/F_{\rm A} = 0.075 \pm 0.012$. This gives an additional indication that the two components have similar spectral classes. Considering the measured flux ratio, this would imply that the secondary spectral class ranges between B2V and B4V.

\begin{figure*}[!t]
       \centering  
       \includegraphics[width=0.85\textwidth]{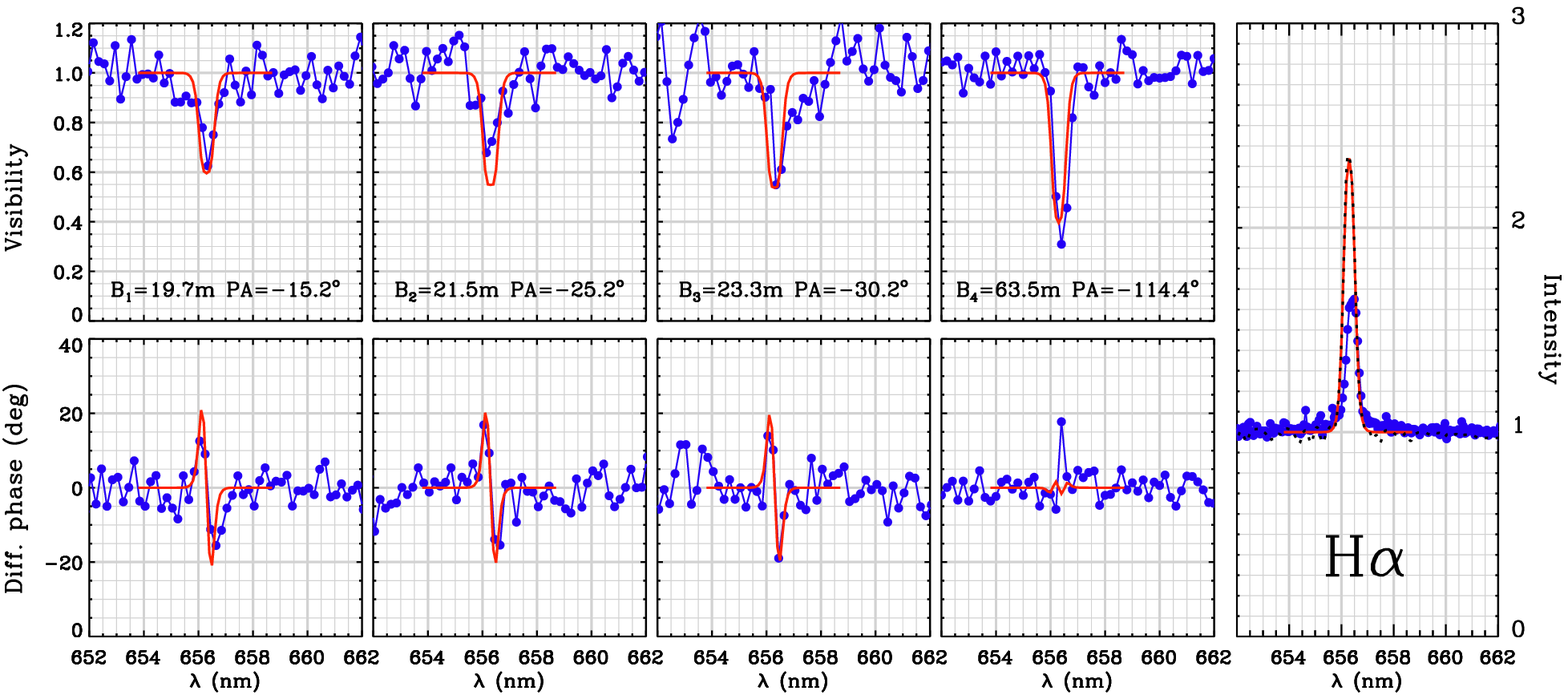}\vspace{0.6cm}
       \includegraphics[width=0.85\textwidth]{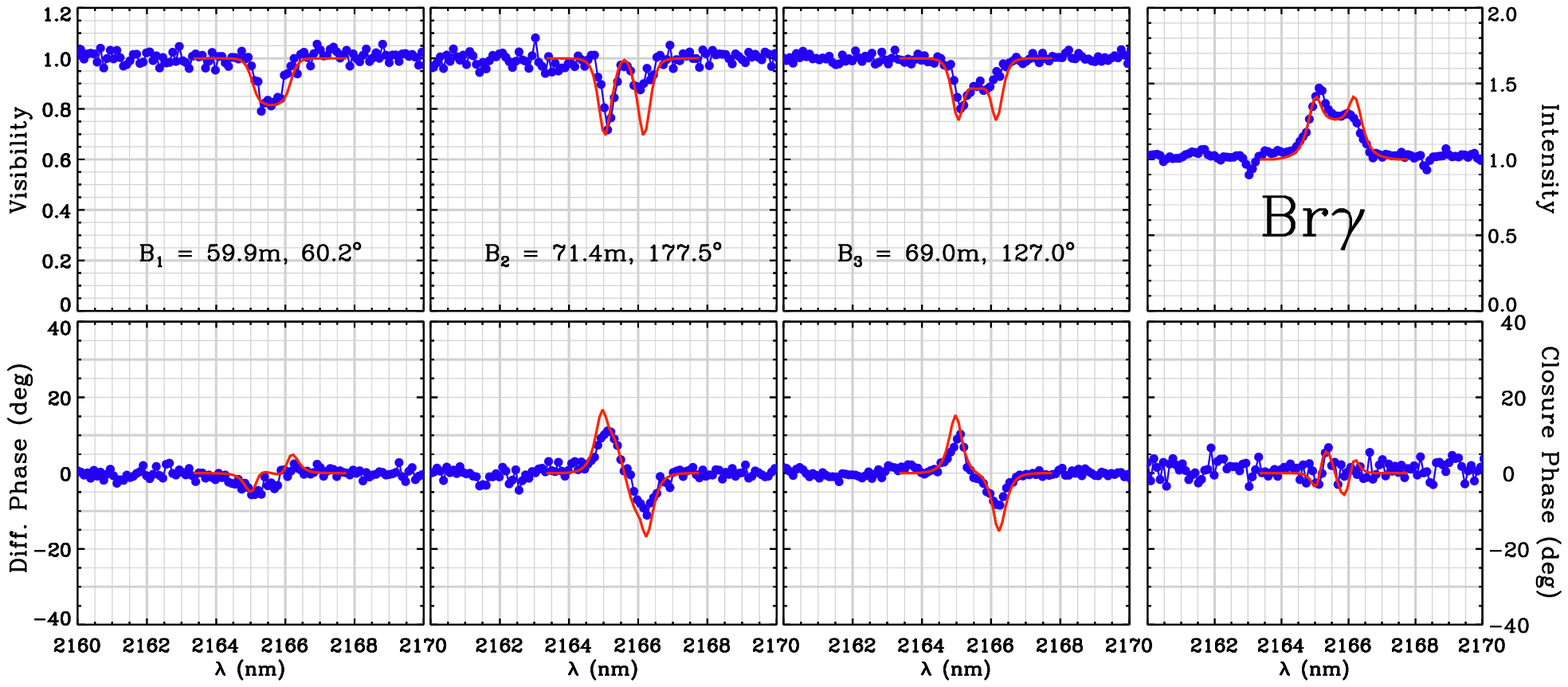}\vspace{0.6cm}
       \includegraphics[width=0.85\textwidth]{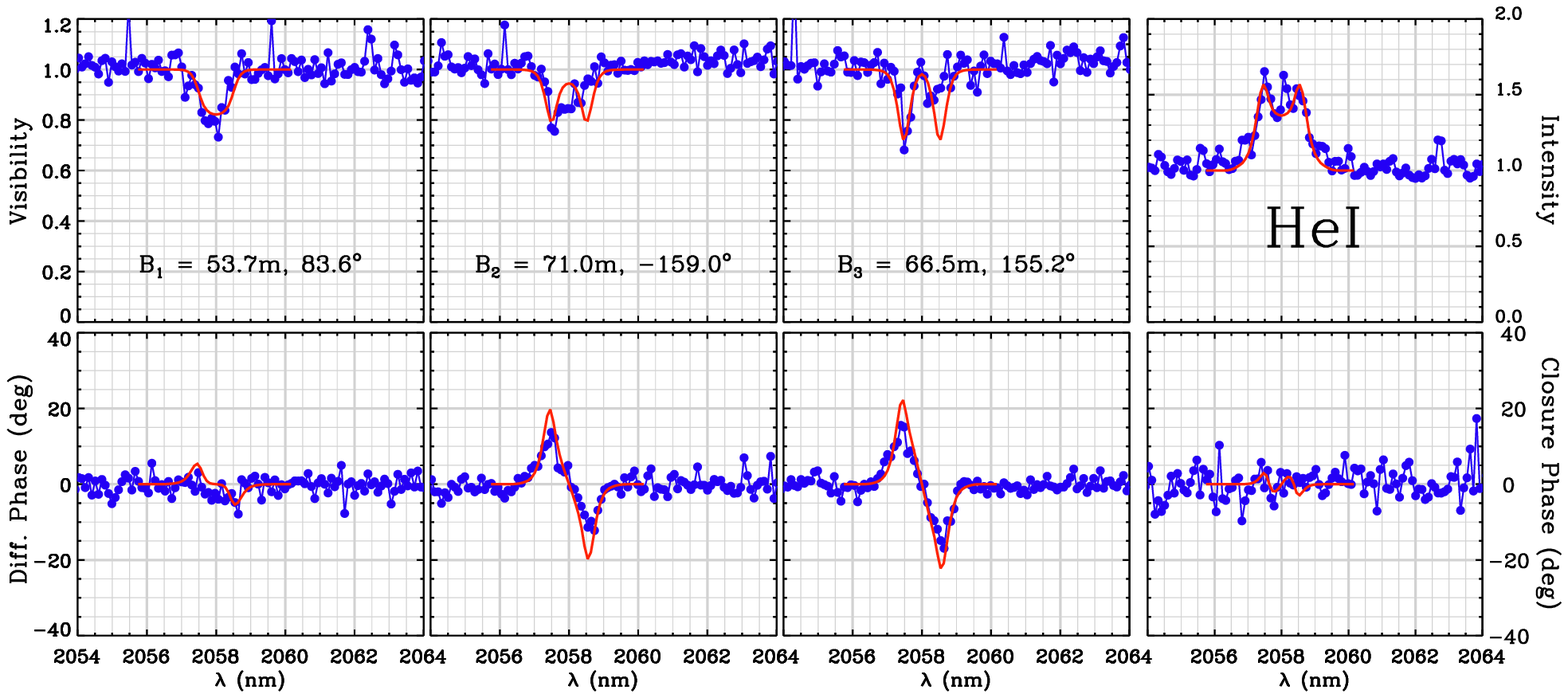}\vspace{0.6cm}
\caption{Visibility and phase variations in the H$\alpha$, Br$\gamma$, and He\,{\sc i} emission lines obtained from the high-spectral-resolution CHARA/VEGA and VLTI/AMBER observations obtained in 2010 (blue line and circles). The best-fit kinematic-model for this dataset is over-plotted as a red line. The black dotted line (upper right figure) corresponds to an H$\alpha$ profile taken from the BeSS database and used to determine the real line EW.}
\label{lines}
\end{figure*}
 
\section{The envelope geometry and kinematics}

During our observing campaign, $\delta$ Sco was observed three times with AMBER in medium resolution (MR): in 2007, 2009, and 2010. All the observations were centered on the fist half of the K band (1.95-2.25\,$\mu$m) in order to study the circumstellar environment geometry and kinematics in the Br$\gamma$ emission line. The quality of the data taken in 2007 was too poor to detect any signal in the line, whereas, as seen in Fig~\ref{visMR}, visibility and phase variations were successfully detected in 2009.  We also discovered that two other K-band emission lines were strong enough to support the kinematics study: the He\,{\sc i} line located at 2.06\,$\mu$m, and the Br$\delta$ line around 1.94\,$\mu$m. However, the latter is located too close to the edge of the K-band so that the data SNR is significantly lower than for the two other lines. We should have decided to study the circumstellar environment geometry and kinematics in MR mode, as it was already done for two other classical Be stars: $\alpha$~Ara (Meilland et al. 2007a) and $\kappa$~CMa (Meilland et al. 2007b). However, since the star was bright enough, we decided to take advantage of the eight times higher spectral resolution offered by the HR mode. With this resolution of $R = 12000$, kinematics details of about 25\,km\,s$^{-1}$ are achievable.

Consequently, in 2010, $\delta$ Sco was observed twice with the VLTI/AMBER in HR mode. The first observations were centered on the Br$\gamma$ line and the second on the He\,{\sc i} line. Unfortunately, we did not manage to calibrate the He\,{\sc i} observation properly. However, this does not affect either the differential visibilities and phases or the closure phase. Finally, quasi-simultaneous medium-spectral resolution observations centered on H$\alpha$ were carried out with the CHARA/VEGA instrument.

\subsection{A qualitative analysis of the high-resolution data}

The differential visibilities and phases, the closure phases (only for VLTI/AMBER data), and line profiles for these observations are presented in Fig~\ref{lines}. Their morphology is similar to those shown by Meilland et al. (2007a) for the Be star $\alpha$ Ara. The ``S'' shape of the phase variation and the ``W'' shape of some visibility variations clearly favor the hypothesis that the circumstellar environment velocity field is dominated by rotation. Using either differential visibility or phase, we can roughly determine the major-axis position angle in the plane of the sky for a purely rotating disk knowing that:

\begin{itemize}
\item The more pronounced ``V'' shape of the visibility i,  the closer the baseline is to the minor axis. The more pronounced ``W'' shape is related to a position closer to the  major-axis. 
\item For the same projected baseline length, the amplitude of the differential phase variation through the emission line is zero if the baseline is aligned with the minor-axis and maximal if aligned with the major-axis. This is true only if the disk is not fully resolved by the interferometer. For larger baselines, the phase amplitude first saturates and then drops, thus second order effects become visible causing the phase to lose its simple ''S'' shape.
\end{itemize}

Thus, both the phase and visibility variations through the observed lines favor the hypothesis of a major-axis roughly in the North-South orientation.

\subsection{The kinematic model}
To model the wavelength dependence of the visibility, the differential phase, and the closure phase in the observed emission lines we use a simple kinematic model developed for fast model fitting of an expanding and/or rotating thin equatorial disk. This model is described in detail in Delaa et al. (2011). The star is modeled as a uniform disk, the envelope emission in the continuum and the emission line has an elliptical Gaussian distribution with a flattening due to a projection effect of the geometrically thin equatorial disk, i.e., $f = 1/cos(i)$, where i the the object inclination angle. The radial and azimuthal velocities are given by:

\begin{eqnarray}
V_{\rm r}&=&V_0 + \left(V_{\infty}-V_0\right)\left(1-\frac{R_{\star}}{r}\right)^{\gamma}\\
V_{\phi}&=&V_{\rm rot}\left(\frac{r}{R_{\star}}\right)^{\beta}~~~~~~~~~~~~~~~~~~~~ 
\end{eqnarray}

For each spectral channel in the line an iso-velocity map projected along the line of sight is then calculated and multiplied by the whole emission map in the line. Finally the whole emission map for each wavelength consists of the weighted sum of the stellar map, the disk continuum map and the emission line map within the spectral channel under consideration. The map is then rotated by the major-axis PA, and scaled using the stellar radius and distance. A 256x256x100 data-cube (i.e. 256x256 for 100 wavelengths) can be computed in less than one second on a standard computer. Finally, visibilities,  differential phases, and closure phases are extracted using two-dimensional fast-Fourier transforms (FFT). We note that the pixel size was set to avoid sampling problems.

The model free-parameters can be classified into 4 categories:  

\begin{enumerate}
\item The global geometric parameters: stellar radius ($R_\star$), distance ($d$), inclination angle ($i$), and disk major-axis position angle ($PA$).
\item The global kinematic parameters: stellar rotational velocity ($V_{\rm rot}$), expansion velocity at the photosphere ($V_0$), terminal velocity ($V_\infty$), exponent of the expansion velocity law ($\gamma$), and exponent of the rotational velocity law ($\beta$). These parameters describe the global disk kinematics and thus also should not depend on the observed line.
\item The disk continuum parameters: disk FWHM in the continuum ($a_{\rm c}$), disk continuum flux normalized by the total continuum flux ($F_{\rm c}$). These parameters depends on the observed wavelength. As the continuum visibility is highly affected by the companion and cannot be constrained properly with our few measurements we decided to neglect this contribution in our modeling. That is why we have chosen to plot differential visibilities (i.e., visibility in each spectral channel divided by the mean visibility) instead of calibrated visibility in Fig~\ref{lines}.
\item The disk emission line parameters: disk FWHM in the line ($a_{\rm l}$) and line equivalent width (EW). These parameters are different for the three emission lines.
\end{enumerate}

Consequently, for simultaneous fit of these three emission lines, we need a model with a total of 15 free parameters. To reduce the number of free parameters we have finally decided:
\begin{itemize}
\item to set the distance to that derived from  Hipparcos (von Leeuween 2007).
\item to use typical values for the expansion velocity law, i.e., $\gamma = 0.86$ and $v_0 = 0$ (Stee \& Araujo 1994; Stee et al. 1995).
\item and to estimate the stellar radius from the fit of the Spectral Energy Distribution (SED). In Carciofi et al. (2006) they  successfully fit the SED using $R_\star = 7$\,$R_\odot$, T$_\mathrm{\rm eff} = 27000$\,K, and $d = 123$\,pc. In Section 3.3,  we found that the distance is better given by the new Hipparcos parallax estimation from van Leeuween (2007), i.e. \textit{d}=~150~pc.  We managed to fit the SED using this distance, the same T$_\mathrm{eff}$ as in Carciofi et al. (2006), and a stellar radius of $R_\star = 150 \div 123 \times 7 \approx 8.5$\,$R_\odot$.
\end{itemize}

Moreover, the lines EW are easily and efficiently constrained by the spectra plotted in Fig~\ref{lines} except for the VEGA/CHARA H$\alpha$ line for which the intensity is underestimated by a factor 2-3 due to a saturation of the photon counting algorithm that is affecting the line EW but not the corresponding visibilities, as already outlined in Delaa et al. (2010). Thus the "true" H$\alpha$ line profile used to compute the EW was taken from the BeSS database\footnote{\tt http://basebe.obspm.fr}  where we found spectra recorded at the same epoch as our interferometric VEGA/CHARA measurements. Finally, running hundreds of models, we tried to constrain the nine remaining free parameters. Values of the best-fit model parameters are presented in Table~\ref{model_params}. The corresponding differential visibilities and phases, closure phases, and lines profiles are overplotted in Fig~\ref{lines}.

\begin{table}[!tbh]
\caption{Values of the best-fit kinematic model parameters.\label{model_params}}
\centering \begin{tabular}{ccc}
\hline
~~~~Parameter~~~~					& ~~~~Value~~~~				&~~~~Remarks~~~~\\
\hline\hline
\multicolumn{3}{c}{\textbf{Global geometric parameters}}\\
$R_\star$				&	8.5						R$_\odot$				& from the fit of the SED\\
$d$							&	150						pc							& from von Leeuween (2007)\\
$i$	  							& 30 			  		deg							& from the fit of the binary\\
$PA$								& -12 $\pm$ 7		deg							&\\
\hline
\multicolumn{3}{c}{\textbf{Global kinematic parameters}}\\
$V_\mathrm{rot}$ 						&	500	$\pm$50		\,km\,s$^{-1}$			&$\approx V_c$\\
$V_\mathrm{0}$ 					&	0							\,km\,s$^{-1}$			& from Stee et al. 1995\\
$V_\infty$					&	0							\,km\,s$^{-1}$			& $<$10m.s$^{-1}$\\
$\gamma$ 					&	0.86													& from Stee et al. 1995\\
$\beta$						&	0.5	$\pm$0.1									& Keplerian rotation\\
\hline
\multicolumn{3}{c}{\textbf{H$\alpha$ disk geometry }}\\
$a_\mathrm{H\alpha}$ 	&	9.0$\pm$3.0		R$_\star$				&= 4.8$\pm$1.5 mas\\
$EW_\mathrm{H\alpha}$	&	7.0$\pm$1.0			$\AA$						&\\
\hline
\multicolumn{3}{c}{\textbf{Br$\gamma$ disk geometry}}\\
$a_\mathrm{Br\gamma}$ 	&	5.5$\pm$1				R$_\star$				& = 2.9$\pm$0.5 mas\\
$EW_\mathrm{Br\gamma}$	&	6.5$\pm$0.5			$\AA$						&\\
\hline
\multicolumn{3}{c}{\textbf{He\,{\sc i} disk geometry}}\\
$a_\mathrm{He\,{\sc i}}$ 			&	4.5$\pm$0.5			R$_\star$				&= 2.4$\pm$0.3 mas\\
$EW_\mathrm{He\,{\sc i}}$		&	8.5$\pm$0.5			$\AA$						&\\
\hline\hline
\end{tabular}

\end{table}
\subsection{The circumstellar disk extension}
We managed to constrain significantly the disk extension in H$\alpha$, Br$\gamma$, and He\,{\sc i}. Modeled as a Gaussian distribution the respective FWHM are $a_{\rm H\alpha} = 9.0 \pm 3.0$\,$R_\star$, 	$a_{\rm Br\gamma} = 5.5 \pm 1$\,$R_\star$, and $a_{\rm HeI} = 4.5 \pm 0.5$\,$R_\star$.	The disk extensions in the infrared lines are compatible with those derived by Millan-Gabet et al. (2010) from 2007 Keck-Interferometer measurements, i.e. $a_{Br\gamma} = 3.6 \pm 0.6$\,R$_\star$ and $a_{\rm HeI} = 4.2 \pm 0.8$\,$R_\star$. Moreover, the size of the H$\alpha$ emission obtained from our CHARA/VEGA measurements is also roughly compatible with the Millan-Gabet et al. (2007) estimation using the line EW, i.e., 14.9\,$R_\star$. Consequently, it seems that the disk did not grow or shrink during the 2007-2010 period. 

However, if we compare our measurements to the physical modeling by Carciofi et al. (2006) from 2005 photometric measurements, i.e. disk outer radius of 7\,$R_\star$, it appears that the disk might have grown between 2005 and 2007 by a factor of at least 1.3. Assuming that the extension of the H$\alpha$ emission represents the physical outer edge of the circumstellar disk, we can estimate the disk expansion velocity. Between 2000 and 2005 the disk has grown from the stellar surface (i.e. 1\,$R_\star$) to 7\,$R_\star$. This represents an average velocity of 0.24\,km\,s$^{-1}$. Between 2005 and 2007 the disk has grown by an additional 2\,$R_\star$, which corresponds to an expansion velocity of 0.19\,km\,s$^{-1}$. These values are the same order of magnitude as those derived for the same object by Miroshnichenko et al. (2003) using the separation of double-peaked emission line profiles assuming Keplerian rotation, and for another variable Be star, Achernar ($\alpha$ Eri), by Kanaan et al. (2008). This value also fully agrees with the model of a viscous decretion disk proposed by Lee et al. (1991) to explain the formation of Keplerian disks around fast rotation Be stars.

Applying a 0.2\,km\,s$^{-1}$ expansion velocity between 2007 and 2010, we should have obtained a disk extension of 12.3\,$R_\star$ in 2010 which is marginaly larger than the extension found in H$\alpha$. However one have to keep in mind that if the disk extends too far from the central star, the density and temperature would drop too low to produce a noticeable H$\alpha$ emission.

\subsection{The circumstellar disk kinematics}
Using our simple kinematic model we managed to put constraints on the disk expansion and rotation velocity laws. As already stated in Section 4.1, the disk kinematics are fully dominated by rotation. An upper limit of 10~\,km\,s$^{-1}$ can be set on the expansion velocity. This is compatible with the value of about 0.2\,km\,s$^{-1}$ derived in the previous section using arguments on the disk evolution. 

Our estimated stellar radius, i.e. $R_\star = 8.5$\,$R_\odot$, is an average photometric radius. Thus, if the star is rotating close to the breakup velocity its equatorial radius will be larger, i.e. on the order of 10\,$R_\odot$, whereas the polar radius will be smaller, i.e. about 7\,$R_\odot$. Using the stellar mass estimated from the binary orbit, i.e. $M_\star = 14$\,$M_\odot$, the stellar breakup velocity is about \textit{V}$_\mathrm{crit} = 500$\,km\,s$^{-1}$.  Consequently, as \textit{V}$_\mathrm{rot} \approx$\,\textit{V}$_\mathrm{crit}$ and $\beta = 0.5$, the  circumstellar matter surrounding the $\delta$ Sco circumstellar disk appears to be in Keplerian rotation. 

\subsection{The disk asymmetry}
To our knowledge, these are the first nearly simultaneously measured spectrally dispersed visibilities with the corresponding differential phases for 3 visible and infrared emission lines, namely H$\alpha$, He\,{\sc i} (2.06\,$\mu$m) and Br$\gamma$. These lines are formed at various distances from the central star as already measured for $\gamma$ Cas in the H$\alpha$, H$\beta$ and He\,{\sc i} 6678 lines by Stee et al. (1998) and as seen in the previous section where we found $\phi_{\rm H\alpha}$ $>$ $\phi_{He\,{\sc i}} > \phi_{Br{\gamma}}$. But compared to Stee et al. (1998) and other recent work (Carciofi et al. 2006; Millan-Gabet et al. 2010) we have now a direct access to the kinematics within each line-forming region with a velocity resolution of about 25\,km\,s$^{-1}$.\\

We see from Fig~\ref{lines} that the differential phases for these 3 lines are, first, well represented by our very simple kinematical model, and second that they are all exhibiting a typical "S" shape of a rotating disk. The amplitude of this "S" shape is on the order of $\pm 10 \degr$ which is very similar to the differential phases obtained by Carciofi et al. (2009) for the Be star $\zeta$ Tau. To the contrary of $\zeta$ Tau differential phases, our differential phases are all symmetrical. $\zeta$ Tau phases showed a very asymmetrical "S" shape with a smaller amplitude in the blue part of the "S" curve with respect to the central line wavelength. This was a clear evidence for a one-armed spiral structure in the $\zeta$ Tau disk.

On the other hand, despite these symmetrical differential phases, our visibilities across these 3 spectral lines are asymmetrical, especially for Br$\gamma$ and He\,{\sc i} lines, with a blue wing of the visibility systematically smaller than the red wing. The line profiles are also clearly asymmetrical, with H$\alpha$ and Br$\gamma$ line profiles showing a \textit{V/R}\,$> 1$ whereas the S/N and spectral resolution for He\,{\sc i} are not high enough to exhibit any line asymmetry.\\

Thus, $\delta$ Sco's circumstellar disk is globally rotating,  with a larger (in size) and brighter emitting region in the blue part of the line, i.e. coming towards us,  and a smaller (more compact), fainter or more absorbed region in the red part of the line, i.e. rotating away from us. These emitting regions are also responsible for the H$\alpha$ and Br$\gamma$ line profiles with \textit{V/R}\,$> 1$. Moreover, since the differential phase is symmetrical, it means that the photocenter of the emitting regions are symmetric with respect to the central star (or the rotational axis).  Therefore, the one-armed oscillation scenario would difficultly match the inhomogeneities detected in $\delta$ Sco's disk since it would have produced non-symmetrical spectrally resolved visibilities {\it and} phases as measured by Carciofi et al. (2009) for $\zeta$ Tau. In our case, the binarity of the system should also play an important role in the shaping and the kinematics of the circumstellar environment. For example, one could think of a tidally-warped disk from previous periastron passages of the companion star (like what is described in Moreno et al. 2011). This definitely needs a dedicated modeling effort to really constrain the detected disk inhomogeneities. 

\section{Conclusion}

Using all available binary separation measurments including our VLTI/AMBER ones, we check the consistency of previous estimates of the $\delta$ Sco  orbital elements. We found parameters that totally agrees with the latest work from Tycner et al. (2011).

Thanks to the combination of high spectral and spatial resolution we managed to constrain for the first time, not only the geometry but also the kinematics of the $\delta$ Sco circumstellar environment. As in the case of Be stars with stable or nearly stable disks such as $\alpha$ Ara (Meilland et al. 2007a), $\kappa$ CMa (Meilland et al. 2007b), $\Psi$ Per, and 48 Per (Delaa et al. 2011), the line emission originates from a dense equatorial disk dominated by rotation. As for most of these objects, the rotation appears to be Keplerian, with an inner boundary (photosphere/disk interface) rotating at the critical velocity (V$\mathrm{c}$). The expansion is negligible and considering an outburst scenario, should be on the order of 0.2\,km\,s$^{-1}$.

Considering the measured vsini, i.e. 175\,km\,s$^{-1}$, and the measured inclination angle, i.e. 30.2\,$\pm$ 0.7$^o$, the star rotates at about 70\,\% of its critical velocity. This could indicate that the stellar rotation is not the main process driving the ejection of matter from the stellar surface. However, taking into account possible underestimation of the vsini due to gravity darkening (Townsend et al. 2004), the star may rotate faster, up to 0.9\,$V_\mathrm{crit}$.

Finally, as the measured extension in the H$\alpha$ line, 4.8\,$\pm$1.5 mas, is on the order of the binary separation at the periastron, i.e. $6.14 \pm 0.07$ mas (i.e., according to Tycner 2011), we expect strong interaction between the previously ejected matter and the companion.

\begin{acknowledgements}
A. Meilland acknowledges financial support from the Max Planck Institut f\"ur Radioastronomy. VEGA is a collaboration between CHARA and OCA/LAOG/CRAL/LESIA that has been supported by the French programs PNPS and ASHRA, by INSU and by the R\'egion PACA. The project has obviously benefitted from strong support of the OCA and CHARA technical teams. The CHARA Array is operated with support from the National Science Foundation through grant AST-0908253, the W. M. Keck Foundation, the NASA Exoplanet Science Institute, and from Georgia State University. This work has made use of the BeSS database, operated at GEPI, Observatoire de Meudon, France: http://basebe.obspm.fr, use of the Jean-Marie Mariotti Center \texttt{SearchCal} service \footnote{Available at http://www.jmmc.fr/searchcal} co-developed by FIZEAU and LAOG, and of CDS Astronomical Databases SIMBAD and VIZIER \footnote{Available at http://cdsweb.u-strasbg.fr/}.This work has also made use of the Jean-Marie Mariotti Center \texttt{LITpro} service co-developed by CRAL, LAOG and FIZEAU. \footnote{LITpro software available at http://www.jmmc.fr/litpro} STR acknowledges partial support from NASA grant NNH09AK731.
\end{acknowledgements}

\end{document}